# On the origin of the optical and near-infrared extragalactic background light

Matsumoto, T.

Department of Space Astronomy and Astrophysics,
Institute of Space and Astronautical Science, Japan Aerospace Exploration Agency,
3-1-1 Yoshinodai, chou-ku, Sagamihara, Kanagawa 252-5210, Japan.

E-mail address: matsumo@ir.isas.jaxa.jp

**Abstract**

In optical and near-infrared background light, excess brightness and fluctuation over the known backgrounds have been reported. To delineate their origin, a fluctuation analysis of the deepest optical images was performed, leading to the detection of a flat fluctuation down to 0.2 arcsec, which is much larger than that expected for galaxies. The sky brightness obtained from the detected fluctuation is a few-times brighter than the integrated light of the galaxies. These findings require some new objects. As a candidate, faint compact objects (FCOs) whose surface number density rapidly increases to the faint end were proposed. FCOs are very compact and show peculiar spectra with infrared excess. If FCOs cause the excess brightness and fluctuation, the surface number density reaches $2.6 \times 10^3$ arcsec$^{-2}$. γ-ray observations require the redshift of FCOs to be less than 0.1 with FCOs consisting of missing baryons. A very low M/L indicates that FCOs are powered by gravitational energy associated with black holes.

## 1. Introduction

Background light refers to the surface brightness of the sky and is important in astronomical observations. Background light is a measure of the total energy in the universe, and its comparison with integrated light of discrete sources determines the extent to which source detections are complete. Background light exists in whole electromagnetic waves, the most well-known of which is the cosmic microwave background (CMB) in radio waves, which provides crucial information about the early universe.

The X-ray background is the first cosmological background identified in the initial phase of X-ray astronomy. Recently, it was found that the faint obscured active galactic nuclei (AGN) could be a probable source of the excess background light.

In the optical and near-infrared regions, the existence of several emission components somewhat complicates the background light, as explained below.

Zodiacal light (ZL) is the most prominent emission component. It is scattered sunlight by interplanetary dust and an important observational target in astronomy. Recent space observations have advanced our understandings on ZL and the interplanetary dust. Using the all sky survey data of Diffuse InfraRed Background Experiment (DIRBE)[1] onboard NASA's infrared astronomy satellite, COsmic Background Experiment (COBE)[2], Kelsall et al. (1998)[3] constructed a physical model of the interplanetary dust cloud by fitting a plausible spatial distribution and optical properties of dust. This model enabled us to estimate the ZL brightness in specific epochs and the wavelength bands of J (1.25 μm), K (2.2 μm), L (3.5 μm), M (4.9 μm) and longer wavelengths. Recently, Matsumoto et al. (2018)[4] reanalyzed the ZL data of Pioneer 10 during the cruising phase to Jupiter and discovered that the spatial distribution of the interplanetary dust is consistent with Kelsall et al.'s model. The spectrum of ZL has been extensively observed, which shows the solar spectrum with a slight enhancement at wavelengths longer than 1 μm.[5]-[8]

ZL is known to be spatially very smooth. The fluctuation of ZL has been estimated based on observations of the thermal emission of the interplanetary dust at mid-infrared wavelengths, where other emission components are negligible.[9],[10] Pyo et al. (2012)[10] could not detect any fluctuation and obtained very low upper limits, i.e. ~0.05% of ZL brightness. Recently, Arendt et al. (2016)[11] analyzed the ZL fluctuation of the near-infrared wavelengths and discovered that the shot noise component of the detected fluctuation at small angular scales was correlated with the ZL intensity.

Unresolved faint stars also contribute to the sky brightness, that is, integrated star light (ISL). ISL depends on the detection limit of each observation, however, it can be directly estimated with the 2MASS star catalog for the J (1.25 μm), H (1.6 μm), and K (2.2 μm) bands, the limiting magnitudes of which are ~15 mag.[12] Due to stars being fainter than the 2MASS detection limits, ISL can be obtained using the model galaxy (cf. TRILERGAL model)[13] with reasonable accuracy. It is now possible to estimate ISL in any direction for a wide wavelength range.

Diffuse galactic light (DGL) is star light scattered by interstellar dust. Based on the DIRBE data, many researchers found that DGL correlates with the far-infrared emission and the DGL fluctuation is proportional to the angular scale.[14]-[17] Recent spectroscopic observations allow the DGL brightness to be estimated more accurately.[18,19]

The integrated light of galaxies (ILG) is emission from unresolved galaxies. Recent deep surveys with large-aperture ground-based telescopes and Hubble Space Telescope (HST)[1] affirm the contribution of faint galaxies to the background.[20]-[22]

Extragalactic background light (EBL) refers to emission that originates from outside the Milky Way. ILG is a constituent of EBL. Our goal is to detect optical and near-infrared EBL accurately while searching for the existence of unknown emission components. For example, the redshifted light of the primeval galaxies and first stars could be observed to be part of the near-infrared EBL,[23]-[27] that provides crucial information on the formation and evolution of the first stars. As far as EBL is concerned, other background emissions such as ZL, ISL, DGL, and ILG are regarded as being foreground emissions.

Although space observations are indispensable for EBL observations due to the strong atmospheric airglow, many researchers encouraged by the challenge to detect the light of the first stars have tried to observe the optical and near-infrared background light by accurately measuring the sky brightness and detecting any spatial fluctuation. The fluctuation measurement has been argued to be advantageous mainly because of ZL's smoothness, which renders its contribution to the sky fluctuation negligible. As for the details of fluctuation measurements and the underlying physics, see the review by Kashlinsky et al. (2018)[28]. Both methods of observation have been tried with a wide range of wavelengths and angular resolutions. The results illustrate the existence of excess brightness and fluctuation in the sky which cannot be explained by known celestial objects.

This paper is structured as follows: Section 2 presents a summary of the previous observations of both the absolute brightness and fluctuation.

---

[1] Details of HST are here; https://hubblesite.org

Section 3 examines the recent analysis of the Hubble eXtreme Deep Field (XDF). Sections 4 and 5 introduce a likely candidate for excess brightness and fluctuation. Section 6 presents a discussion and Section 7 presents the conclusion.

2. Previous observations of the optical and near-infrared EBL
2.1 Observations of the absolute sky brightness and EBL

In the near-infrared region, sounding rockets were used to perform the first space observations.[29)-31)] These observations succeeded to accurately measure the absolute near-infrared sky brightness; however, they were unable to separate the brightness to components well, mainly because the observed sky was limited due to the brief observation time. Afterward, observations with satellite-borne telescopes became popular and managed to provide useful data, which are summarized in Fig. 1.

DIRBE observed 90% of the entire sky from near-infrared to far-infrared wavelengths. Based on the DIRBE/COBE archive data and the model, many researchers have detected excess near-infrared EBL over ILG at the J (1.25 µm), K (2.2 µm) and L (3.5 µm) bands.[32)-39)] Some of them used a slightly different model ZL[40)]; however, Matsumoto et al. (2015)[41)] showed that both models render consistent results within the errors. Recently, Sano et al. (2015[42)],2016[43)]) reanalyzed the DIRBE/COBE data using the latest observational results of ISL and DGL. They managed to obtain more reliable near-infrared EBL, consistent with the previous results. Fig. 1 shows their results as representative findings of COBE observations (black filled squares).

The Near-InfraRed Spectrometer (NRS)[44)], one of the instruments onboard the InfraRed Telescope in Space (IRTS)[45)], succeeded to observe diffuse background light at 1.4 ~ 4 µm using low-resolution spectroscopy. Using the NIRS/IRTS data, Matsumoto et al. (2005)[46)] reported on the detection of excess brightness, which can hardly be explained with the known sources. Fig. 1 shows the recent reanalysis with the latest observational results on ISL and DGL.[41)] The obtained near-infrared EBL is well consistent with that obtained with DIRBE/COBE in the common wavelength region ($\lambda > 2$µm). Matsumoto et al. (2005)[46)] also analyzed the sky fluctuation, and found excellent correlations between the wavelength bands.

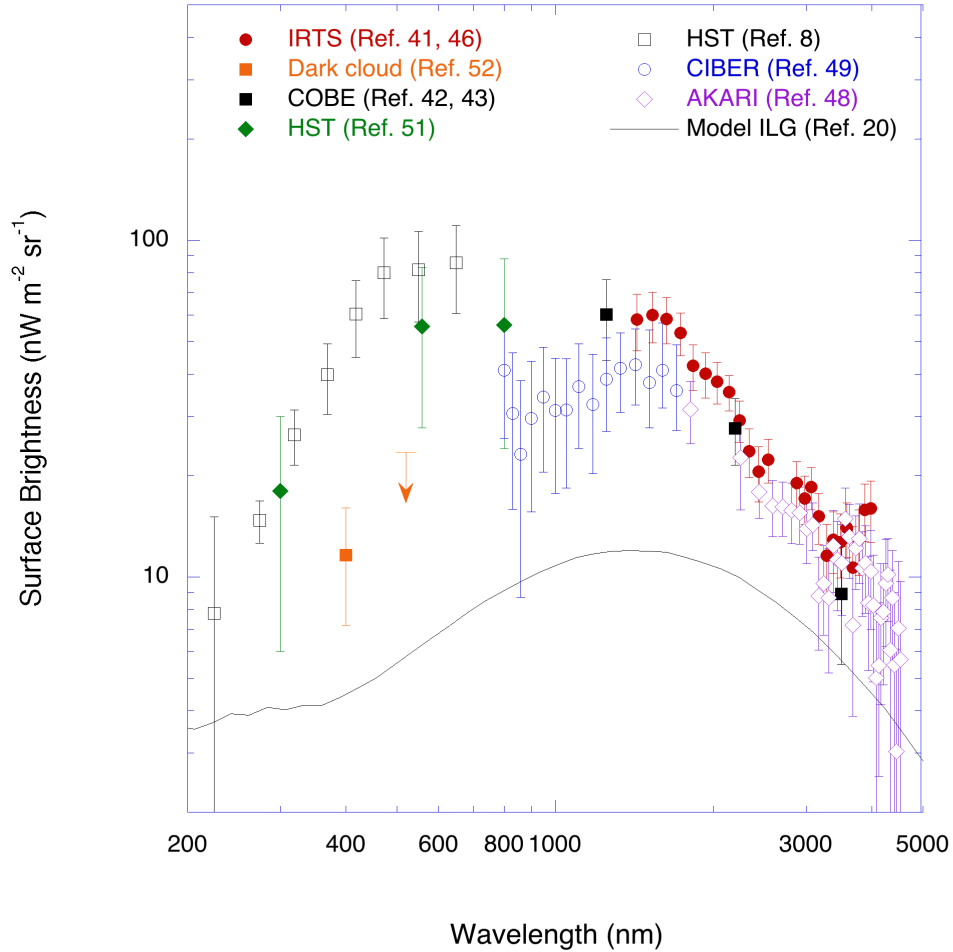

Fig. 1 Summary of previous observations of EBL. Error bars indicate 1σ, quoted from the original papers. The solid black line indicates model ILG.

Using a Japanese infrared astronomy satellite, AKARI[47], Tsumura et al. (2013)[48] detected EBL at wavelengths longer than 3 μm (see purple open diamonds in Fig. 1), which echoes the results of COBE and IRTS. Consistency in three distinctly different observations reveals the existence of unknown isotropic emission at wavelengths longer than 1.4 μm.

The spectra observed with COBE, IRTS and AKARI show a stellar feature whose peak locates in a shorter wavelength region. Some researchers suspect that the reported EBL has a ZL origin, since the spectrum is similar to that of ZL.[38] To confirm this, Matsuura et al. (2017)[49] attained a sounding rocket experiment, Cosmic Infrared Background ExpeRiment (CIBER)[50], to observe EBL at around 1 μm with low-resolution spectroscopy. After subtracting any foreground emission components, they detected EBL whose spectrum is fairly different from ZL (blue open circles in Fig. 1). Furthermore, they did not detect any spectral features of redshifted Lyman α and/or Lyman limit, which are expected from the theory of first stars[38].

Even in the optical wavelength region, EBL has been detected, although foreground emissions are more serious than those in the near-infrared region. Bernstein (2007)[51] observed the optical sky brightness using the Wide Field and Planetary Camera 2 (WFPC2) on HST and subtracted ZL by observing Fraunhofer lines with a ground-based telescope. The results (green filled diamonds in Fig. 1) indicate a high level of EBL, despite fairly large error bars.

Using Faint Object Spectrograph (FOS), Kawara et al. (2017)[8] analyzed the spectra of the blank sky, and detected bright isotropic residual emission (see black open squares in Fig. 1) which is consistent with Bernstein's observation[51]. They speculate that the residual emission may be isotropic ZL, mainly because of the similarity of the spectrum of the residual emission to that of ZL. However, it is extremely difficult to assume such an isotropic emission, because the interplanetary dust and solar radiation field have a heliocentric distribution, whereas the observed isotropic emission indicates geocentric features.

Using a very large ground-based telescope, VLT, Mattila et al. (2017)[52] studied the spectroscopy of the stellar Fraunhofer lines in dark clouds at high galactic latitudes to estimate EBL by comparing the line's strength inside and outside the dark cloud. The authors reported that they managed to detect EBL at 400nm with an upper limit at 520 nm (see the orange filled square and the bar in Fig. 1). They also noticed that the detected EBL was significantly higher than ILG.

The observation of EBL from outside the zodiacal cloud is very attractive, mainly because ZL is the most prominent foreground and the major source of uncertainty. Following Toller's (1983)[53] pioneering work, Matsuoka et al. (2011)[54] analyzed the ZL data of Pioneer 10 and 11 during the cruising phase and traced the optical sky brightness up to Jovian orbit. They claimed that the sky brightness near Jupiter's orbit was consistent with ILG. However, a recent reanalysis confirmed the existence of instrumental offsets, which renders the observation of absolute sky brightness impossible[4]; hence, we did not plot Matsuoka et al.'s results[54] in Fig. 1.

The solid black line indicates the ILG model of the ordinary galaxies[20], which is consistent with the recent deep galaxy surveys. The estimated ILG errors in these wavelength ranges are less than 20%, ~ 2 nWm$^{-2}$sr$^{-1}$,[22] which are negligible compared with the detected EBL brightness. The observed optical and near-infrared EBL shows significantly higher brightness than ILG, which suggests the existence of unknown emission sources.

2.2. Observations of the fluctuation of the sky

As noted earlier, measuring the fluctuation of the sky is advantageous because it can reduce the contribution of ZL arising from its smoothness.[9,10] Following the pioneering work on the fluctuation study of the 2MASS filed[55], Kashlinsky et al. (2005)[56] succeeded in detecting the fluctuation of the source subtracted images obtained from Spitzer Space Telescope[57] at 3.6, and 4.5 μm. The researchers discovered excess fluctuation which can scarcely be explained using ordinary galaxies. They confirmed excess fluctuation in different sky regions.[58,59] Matsumoto et al. (2011)[60] and Seo et al. (2015)[61] performed a similar fluctuation analysis in the north ecliptic pole (NEP) using the Japanese infrared astronomy satellite, AKARI images at 2.4, 3.2 and 4.1 μm. These researchers observed similar excess fluctuation at angles larger than 100 arcsec. They also discovered a strong correlation between the wavelength bands, and noticed that the wavelength dependence of the fluctuation was consistent with the Rayleigh-Jeans law. However, Helgason and Komatsu (2017)[62] have suggested that the detected fluctuations may be explained by clustering of ordinary galaxies.

HST images have been used too for fluctuation analysis. Thompson et al. (2007)[63,64] and Donnerstein (2015)[65] detected an excess fluctuation for the images of the NICMOS deep field (NDF) at 1.0 and 1.6 μm. Mitchell-Wynne et al. (2015)[66] analyzed the optical and near-infrared images of the Cosmic Assembly Near-infrared Deep Extragalactic Legacy Survey (CANDELS)[67] field and obtained excess fluctuation and a good correlation between the wavelength bands. Zemcov et al. (2014)[68] performed a sounding rocket experiment; CIBER, and detected excess fluctuation at around ~ 10 arcmin scale at 1.0 and 1.6 μm.

Recently, Kim et al. (2019)[69] constructed two-dimensional images of the IRTS data in the H and K bands and found that excess fluctuation extends to degree scales and turns over beyond one degree.

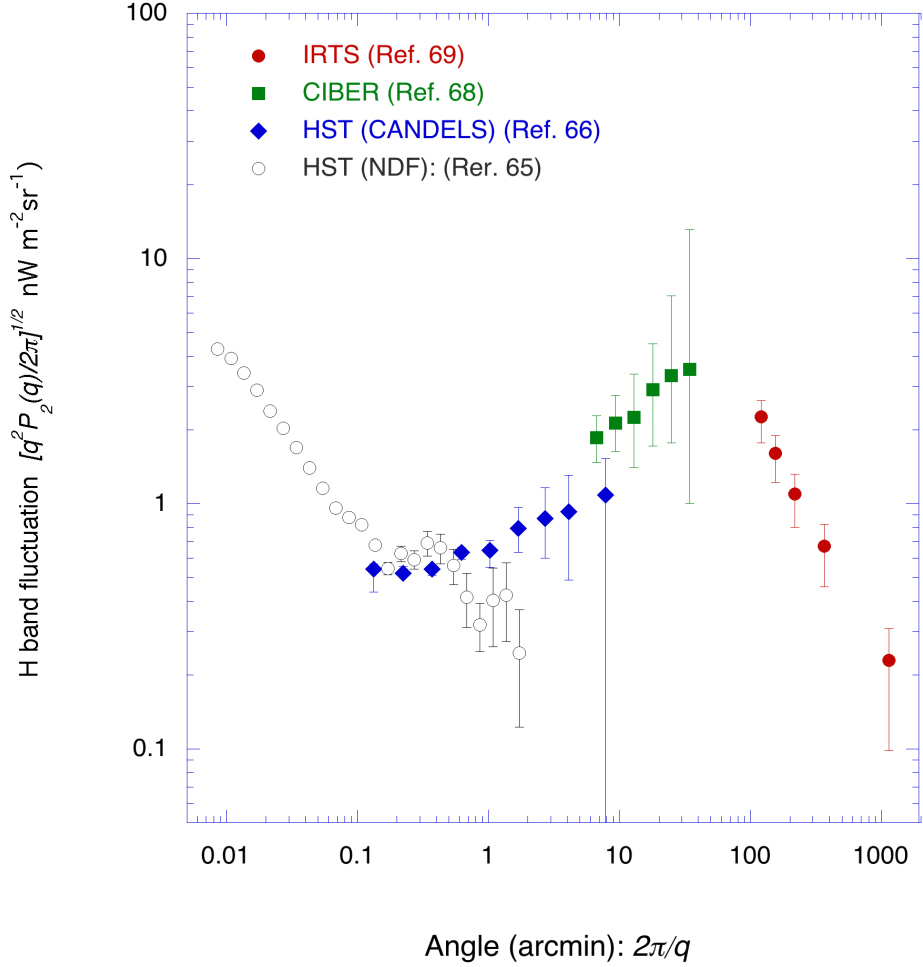

Fig. 2 Angular dependence of the fluctuation of the sky in the H band. Error bars indicate 1σ, quoted from the original papers.

Fig. 2 shows the angular dependence of the observed fluctuations in the H band. Sources fainter than ~12, ~15, ~27, and ~28 AB magnitudes are masked for images of IRTS[69], CIBER[68], HST (CANDELS)[66] and HST (NDF)[65], respectively. The ordinate indicates fluctuation, $(q^2 P_2(q)/2\pi)^{1/2}$, where $P_2(q)$ and $q$ represent the power spectrum and wavenumber, respectively. A steep rise at small angles is due to the shot noise of faint galaxies. A large fluctuation can be seen at angles of 1 ~ 100 arcmin, though it decreases at degree scales. Shot noise and flat fluctuation are commonly observed at different wavelengths.

Another important finding is the excellent correlation between the near-infrared EBL and the unresolved X-ray backgrounds. Cappelluti et al. (2013)[70] and Cappelluti et al. (2017)[71] noticed a good cross-correlation in the fluctuation between the Spitzer 3.6 and 4.5 μm bands, and Chandra[72] X-ray bands: [0.5–2] keV. The correlation with harder X-ray bands: [2–7] keV was found to be marginally significant. The authors attributed the detected correlation at an angular scale of smaller than 20 arcsec to known sources, such as AGN, starburst galaxies, and hot gas, however, residual excess fluctuations remain at large angular scales. Since the X-ray background has a peak near 30 keV, the X-ray sources responsible for the residual fluctuation could be new unknown objects.[71,73]

2.3 Origin of excess brightness and fluctuation

Numerous researchers have examined the origin of the excess brightness and fluctuation. First stars are an attractive candidate; however, recent evolution models based on the observations of high redshift galaxies indicate that the brightness and fluctuation caused by first stars are much lower than the observed excesses.[73-76]

Yue et al. (2013)[77] explained the fluctuation observed with the Spitzer Space Telescope using direct collapsed black holes (DCBHs) at high redshifts. Helgason et al. (2016)[78] examined the possibility of using emissions from black hole remnants of massive first stars to explain near-infrared excess fluctuation. Kashlinsky (2016)[79] examined near-infrared fluctuation due to the primordial black holes detected by LIGO[80] and found primordial black holes (PBHs), of the LIGO-mass range, need to make up the dark matter in order to explain the observed near-infrared excess fluctuations. A black hole is a favorable object for explaining the cross-correlation between the near-infrared and X-ray backgrounds.[70,71] Despite this, a serious difficulty hinders explaining the observed excess brightness in the optical wavelength band, because the redshifted Lyman α and Lyman limit must exist at $\lambda > 1$ μm. Ricarte et al. (2019)[81] conducted simulation studies on the fluctuation of DCBH at redshift of several, but failed to obtain a reasonable fit.

Cooray et al. (2012)[82] proposed intra-halo light (IHL). IHL refers to emission that is released from stars expelled into the intergalactic space due to collisions and/or the merging of galaxies at low redshifts. IHL is useful in energetics, although no observational evidence of IHL has been reported to date. However, there are difficulties to invoke IHL to explain large fluctuations at the degree scale[69] and the excess EBL. Furthermore,

IHL cannot explain a good correlation between near-infrared EBL and the X-ray background.

At this stage, no known source can explain all of the EBL observations, meaning that new emission sources are needed.

3. Fluctuation analysis of Hubble eXtreme Deep Field (XDF)

To identify the origin of excess brightness and fluctuation, Matsumoto and Tsumura (2019)[83] (hereafter referred to as M&T) analyzed the fluctuation of XDF, which is the deepest image using HST.[84] They expected the appearance of new objects to be responsible for the excess brightness and fluctuation, since ILG arising from the ordinary galaxies is very low in XDF images after masking.

The area of the XDF is 11 arcmin$^2$. However, M&T analyzed the central part with a 60 milli-arcsec (mas) pixel resolution where the integration time is the longest and most uniform. They analyzed optical images whose filter names and central bandwidths were F435W (432.3 nm), F606W (591.2 nm), F775W (769.9 nm), and F850LP (906.0 nm), respectively.

For a fluctuation analysis, M&T constructed a mask image by masking all plausible sources in a combined image (see details in Ref. 83), which was applied to the individual images. Eventually, 53.3% of the pixels remained. Fig. 3 shows the extracted images at the F775W band, where the left and right panels correspond to the original and masked images, respectively. The masked image illustrates the existence of a large-scale spatial structure. The limiting magnitudes of the masked images are defined as 2 σ for 2 x 2 pixels resolution, which amounts to 33.19, 33.62, 33.36, and 32.80 in AB magnitude for F435W, F606W, F775W, and F850LP, respectively.

For the masked image, M&T performed a two-dimensional Fourier transformation and obtained the power spectrum, $P_2(q)$, fluctuation, $(q^2 P_2(q)/2\pi)^{1/2}$, as a function of the wavenumber, $q$, following previously reported analyses[63]. They also applied a simple correction for the masking effect, which is only slightly noticeable at large angles.[83]

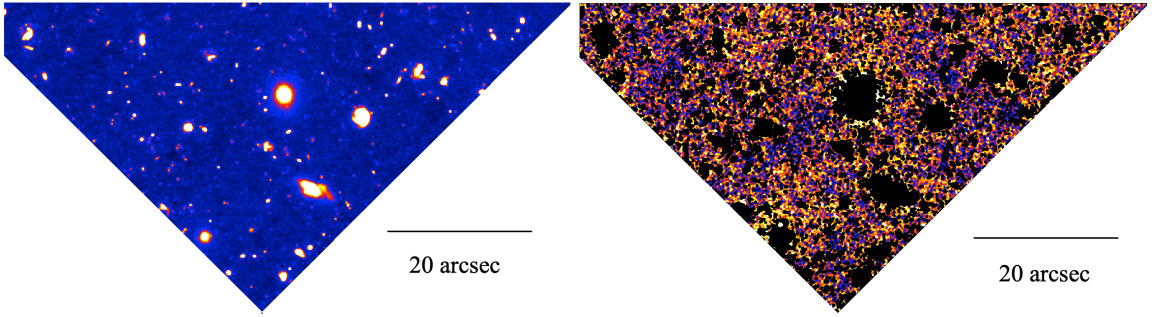

Fig. 3 Retrieved XDF images at the F775W band used for fluctuation analysis. The left and right panels correspond to the original and masked images, respectively.

Fig. 4 presents the fluctuations of four filter bands as a function of the angles, $2\pi/q$. The shot noise at small angular scales and flat fluctuation patterns at angles larger than 10 arcsec are evident. A flat fluctuation corresponds to $1/f$ distribution: it implies a fractal structure, which is commonly found in nature. Black lines represent the fluctuations expected of galaxies fainter than the limiting magnitudes, which are estimated based on the surface number density magnitude relation in XDF field (see black lines in the following Fig. 9) and evolution model[85]. The detected fluctuation levels are much higher than those expected of the ordinary galaxies (black lines in Fig. 4).

  M&T performed a cross-correlation. Fig. 5 shows both auto-fluctuations and cross-fluctuations for all combinations. The shot noises remain for the cross-correlations with F435W, which could be due to the remaining ordinary galaxies because the detection limit of the F435W band is fairly worse than the other three bands. The shot noise disappears for those cross-correlations with the three other bands, indicating that the shot noises at small angular scales can be ascribed to photon noise arising from bright ZL. The cross-correlations among the three longer wavelength bands are excellent.

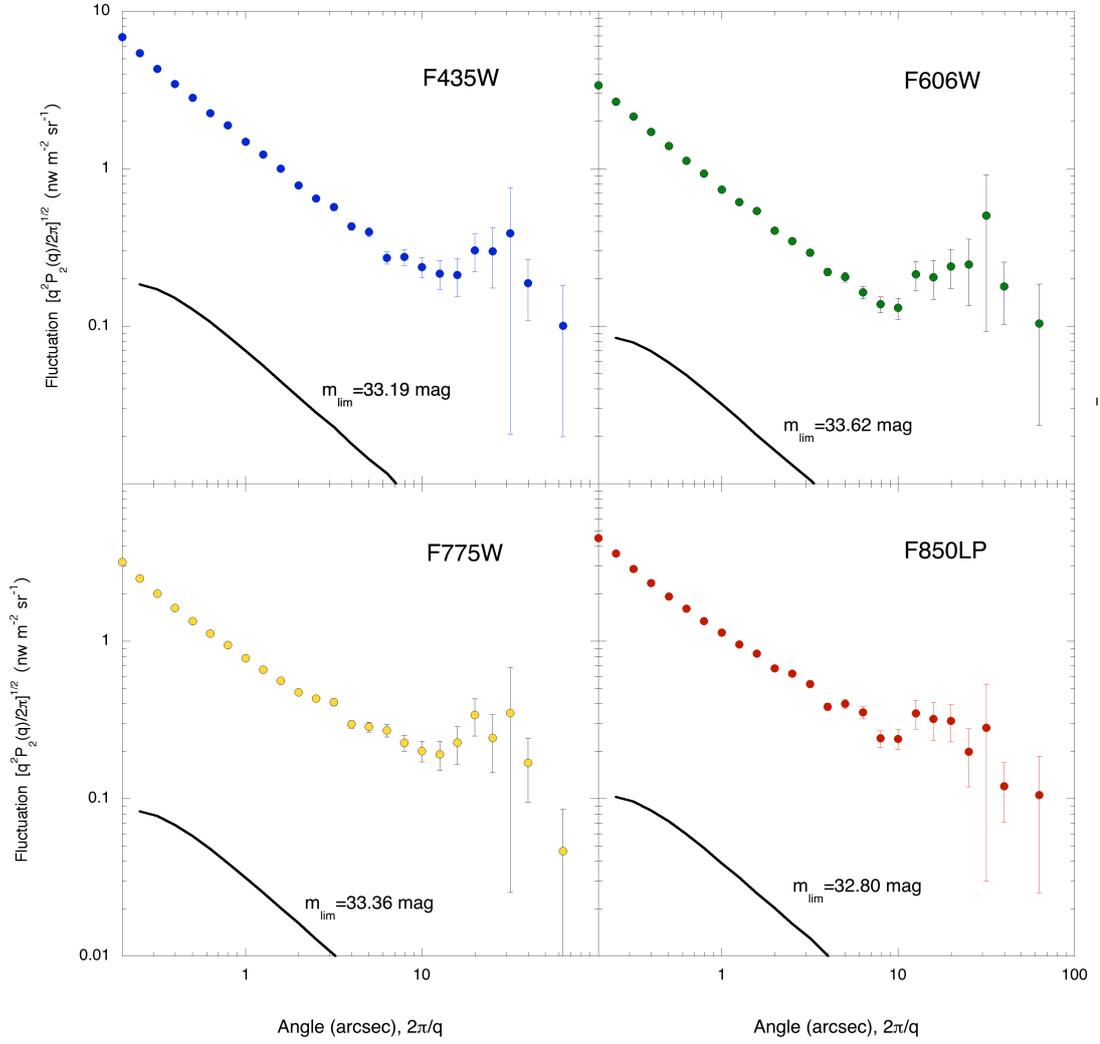

Fig. 4 Auto-fluctuation for four wavelength bands. Black solid lines indicate the fluctuation of ILG from ordinary galaxies fainter than $m_{lim}$.

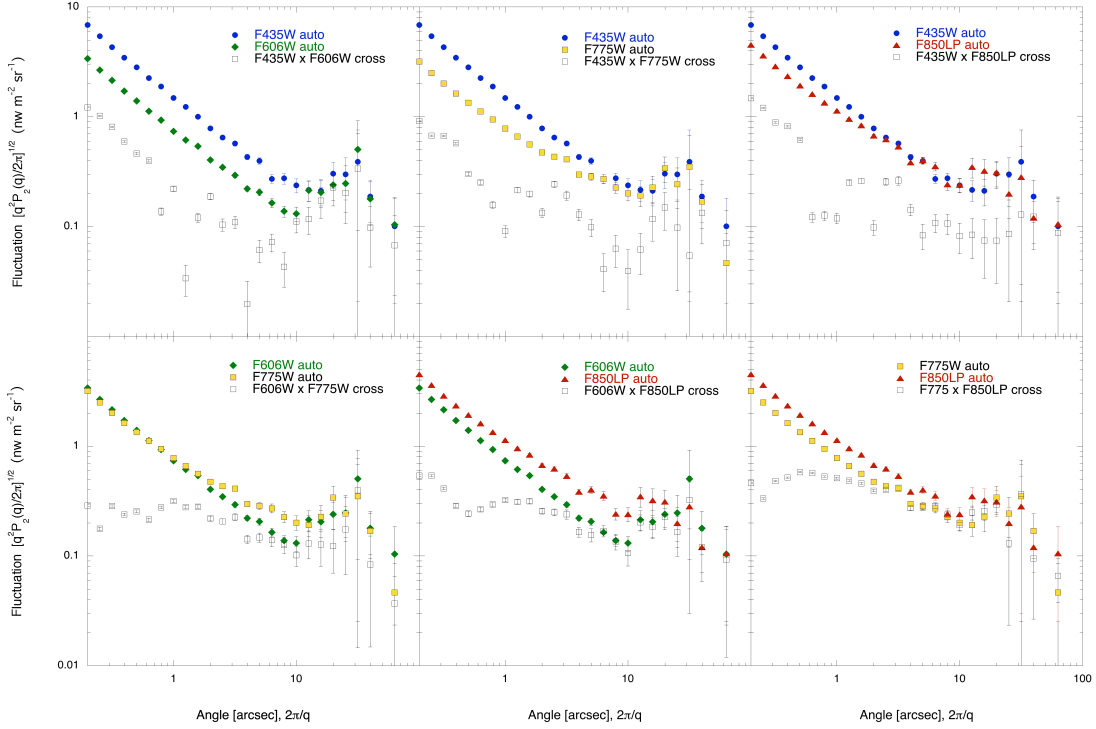

Fig. 5 Cross-fluctuations for six combinations are shown in the open black squares together with two corresponding auto-fluctuations that are the same as those in Fig. 4.

Fluctuations of the cross-correlations are marginally flat from 0.2 to 100 arcsec, indicating the existence of sky structure, even at the 0.2 arcsec scale. The fluctuation is probably caused by source confusion and a specific spatial distribution of unknown sources. In other words, the sky is filled with unknown objects, even on the 0.2 arcsec scale. The surface number density for one object in a 0.2 arcsecond square corresponds to $3.2 \times 10^8$ degree$^{-2}$. This value is a lower limit, but is much larger than that of ordinary galaxies, as is discussed in the next section.

Large correlations imply that considerable sky brightness is due to unknown objects. The auto-fluctuation spectra of the F775W and F850LP bands must be larger than the cross-fluctuation spectrum, F775W × F850LP, since there may be an additional uncorrelated fluctuation component in each band. Therefore, M&T assumed that the cross-fluctuation spectrum is the minimum of the EBL auto-fluctuation spectra for both the F775W and F850LP bands. They created simulated images with a fluctuation spectrum that is same as the cross-fluctuation spectrum (F775W × F850LP), and obtained a dispersion of 1$\sigma$ for the brightness distribution. Using 100 trials, they obtained a 1 $\sigma$ value of 8.10 ± 0.01 nWm$^{-2}$sr$^{-1}$. Since the sky brightness must always be positive, M&T took the 3 $\sigma$ brightness, 24 nWm$^{-2}$sr$^{-1}$, as the lower limit of the absolute sky brightness in the F775W and F850LP bands.

M&T estimated the ILG emanating from ordinary galaxies fainter than the limiting magnitudes based on the surface number density magnitude relation (see black lines in the following Fig. 8). They obtained 0.053 and 0.056 nWm$^{-2}$sr$^{-1}$ for F775W and F850LP bands, respectively, which are negligible compared with the obtained lower limit. Furthermore, the detected lower limit is so bright that the IHL origin for excess brightness and fluctuation is rejected.[82]

Fig. 6 shows excess brightness after subtracting the model ILG in Fig. 1, since the observations all have different limiting magnitudes. The uncertainty of ILG is at most 2 nWm$^{-2}$sr$^{-1}$,[22] which is negligible compared with EBL observation errors. The obtained lower limit is marked by a red thick horizontal bar, which covers the wavelength range of the F775W and F850LP bands. Fig. 6 shows that M&T's findings are consistent with those reported in previous studies. It must be emphasized that the obtained excess brightness is a lower limit. If the flat fluctuation spectra extend to degree scales similar to the H band in Fig. 2, the excess brightness could be a few-times brighter, which is comparable with Kawara et al.'s (2017)[8] and Bernstein's (2007)[51] result.

The importance of M&T's work is that excess EBL is confirmed to be free from the uncertainty in the ZL subtraction. Arendt et al. (2016)[11] analyzed the sky fluctuation in near-infrared wavelength bands at small angles and detected the shot noise component due to the photon noise of ZL, which could be part of the sky fluctuation at angles of less than 10 arcsec in Fig. 4. Even if the ZL fluctuation has some structures, its fluctuation pattern is not fixed on the sky, and must show a temporal variation corresponding to the revolution of Earth and dust clouds around the Sun. Since the observations at each wavelength band were performed in the different epochs, the excellent cross-correlations in Fig. 5 cannot have a ZL origin.

In summary, M&T revealed the existence of bright excess background light, even at optical wavelengths. They also showed that the energy distribution in the optical and near-infrared wavelengths is a few-times higher than that of ILG. These indications illustrate the presence of unknown celestial objects, which are densely distributed across the sky.

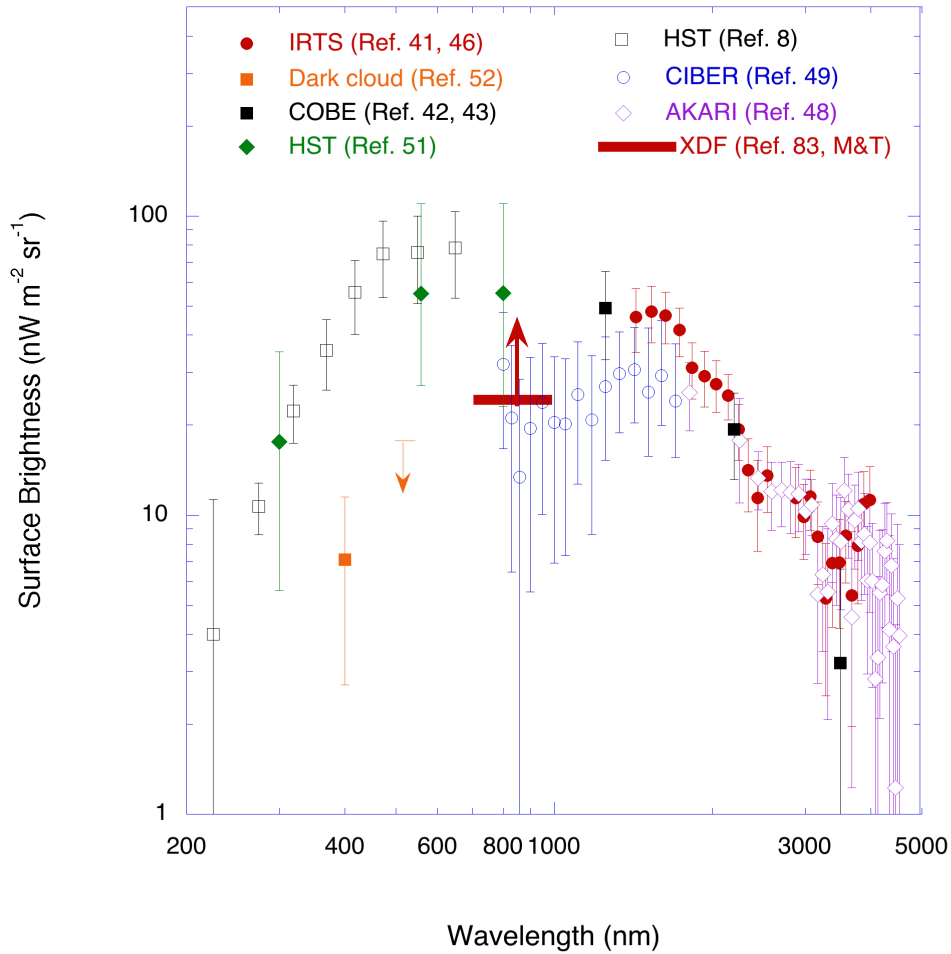

Fig. 6 Summary of excess brightness over ILG of ordinary galaxies. The thick red bar indicates the M&T's lower limit.

4. Faint compact objects (FCOs) in XDF

To identify the unknown objects, M&T examined XDF images and found that the density of compact objects dramatically increases toward the faint end. Given the absence of an XDF catalog, M&T used Beckwith et al.'s (2006)[86] Ultra Deep Field (UDF) catalog to investigate the properties of 8918 objects in XDF images.

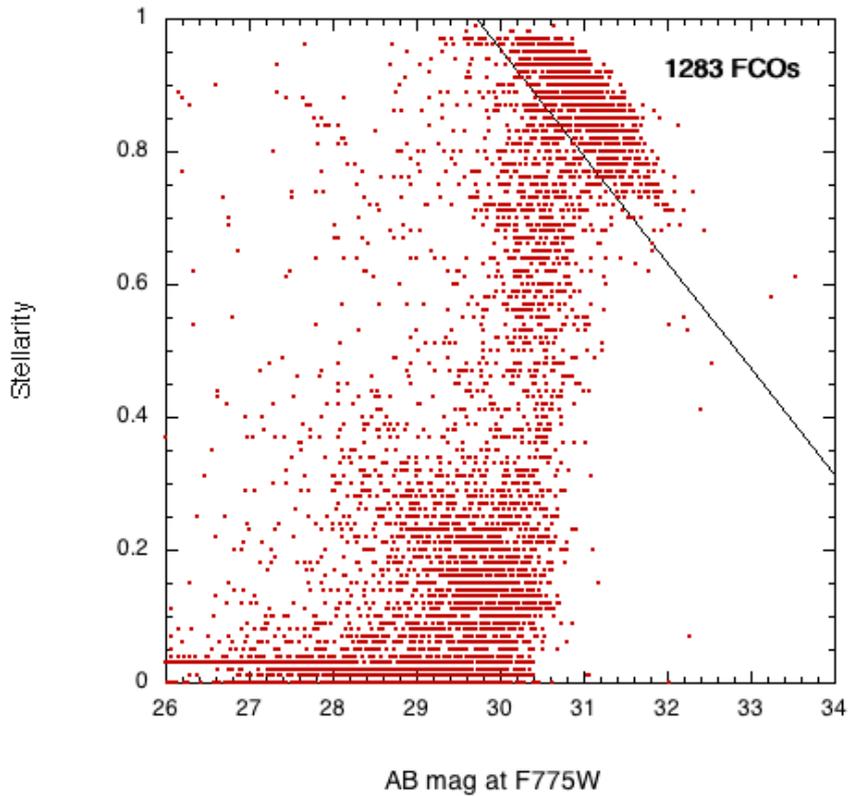

Fig. 7 Stellarity magnitude relationship of UDF objects for the F775W band.

Fig. 7 shows the dependence of stellarity on the AB magnitude of the F775W band, because stellarity indices in the catalog were measured for the F775 band. The stellarity index was developed to identify the stars in images by evaluating the compactness and presence of an extended envelope.[87] Objects with a stellarity index of larger than 0.8 are usually perceived as stars. Candidates of stars in our Galaxy[88] are removed, although their magnitudes are brighter than 26 AB mag, and the numbers are negligibly small.

A somewhat arbitrary line is drawn in Fig. 7 to separate FCOs from ordinary galaxies. The number of FCOs, the objects located toward the upper right of the solid line, is 1283. For a detailed analysis, M&T retrieved 649 FCOs whose stellarities are larger than 0.9.

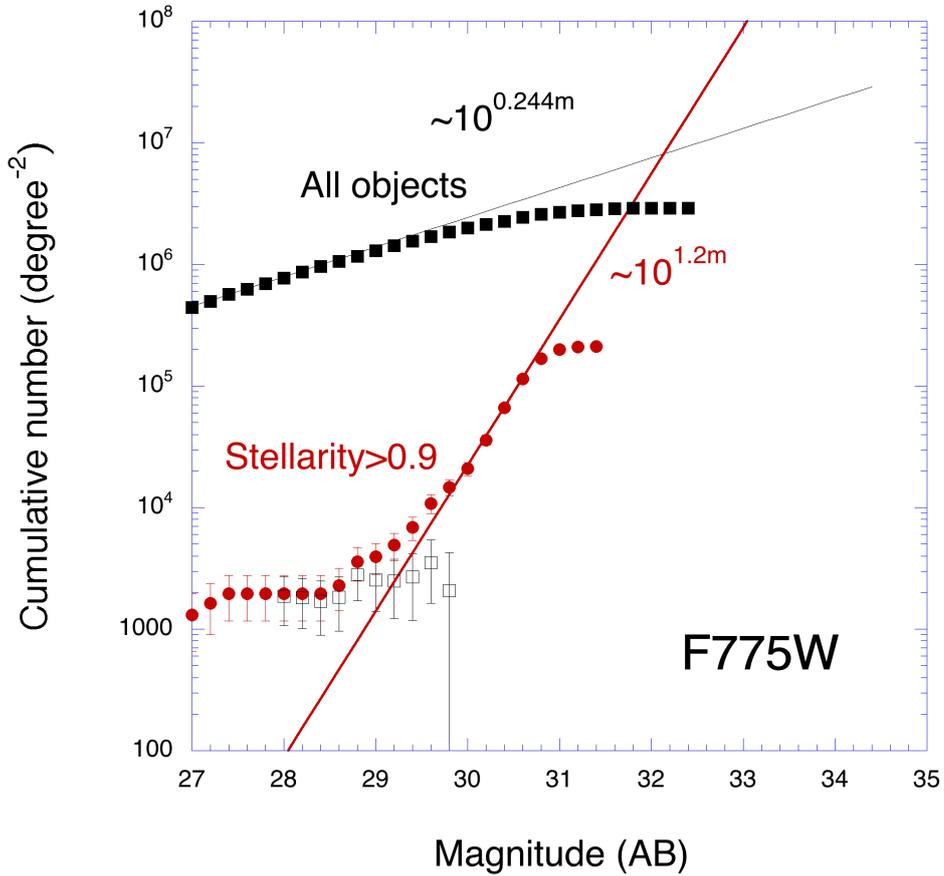

Fig. 8 Dependences of the cumulative number densities on the AB magnitude of F775W are shown for all objects (black filled squares) and FCOs (red filled circles), respectively.

Fig. 8 indicates the cumulative number counts as a function of the apparent AB magnitude, $m$, for the F775W band. The black filled squares and red filled circles represent all objects and objects with stellarity $> 0.9$, respectively. Fig. 8 shows that the surface number density of FCOs increases much more dramatically than that of all objects toward the faint end. M&T fits the slope for FCOs using a power law (red lines) such that the residual numbers of objects in the bright region are constant (black open squares). The result, $\propto 10^{1.2m}$, is much steeper than the isotropic distribution, $\propto 10^{0.4m}$, and corresponds to the density distribution, $n(r) \propto r^3$, where $n(r)$ signifies the spatial number density and $r$ represents the radial distance. The surface number density of FCOs must turn over at a certain magnitude; otherwise, the sky brightness would diverge.

By applying the evolution model[85], M&T estimated that the slope of all objects toward the faint region was $\sim 10^{0.244m}$. Since the model predicts a smooth slope up to 32 AB mag, the turnovers of all the detected objects and FCOs are due to the detection limits. If we assume a simple extrapolation and consider the total number of FCOs to be twice the value given in Fig. 8, the surface number densities of FCOs exceed those of ordinary galaxies at 31.8 AB mag. If we assume that the excess surface brightness is due to the integrated light of FCOs, the lower limit of the surface brightness of 24 nWm$^{-2}$sr$^{-1}$ implies a cutoff magnitude of 34.9 AB mag for the F775W band. In this case, the cumulative surface number density is determined to be $3.35 \times 10^{10}$ degree$^{-2}$, $2.6 \times 10^{3}$ arcsec$^{-2}$, or $1.38 \times 10^{15}$ for the entire sky.

To confirm the validity of the stellarity indices, M&T examined the compactness of bright FCOs using images with a pixel resolution of 30 mas. They divided the images into annular rings with a width of 0.03 arcsec and obtained the average surface brightness in units of $\lambda I_\lambda$: nWm$^{-2}$sr$^{-1}$ for each ring. Using the same procedure, they obtained radial profiles of the point spread function (PSF) using stellar images for comparisons with those of FCOs. Fig. 9 shows a typical example of the radial profiles for a bright FCO, UDF8942, which are not distinguishable from the stellar images of all wavelength bands. Furthermore, no extended envelope has been detected for any of the wavelength bands. They conducted a similar analysis for bright FCOs, and confirmed that all of them show undistinguishable radial profiles from PSF, except for the five bright objects earlier identified as galaxies.

To quantitatively estimate the compactness of FCOs, M&T used de Vaucouleur's law, which represents the radial profile of the dwarf spheroids:

$$I(r) = I_e \exp\{-7.67[(r/r_e)^{1/4} - 1]\},$$

where $r_e$ is the effective radius in which half of the total light is contained. By fitting the model with different $r_e$ values, M&T obtained the upper limit of the effective radius to be 0.02 arcsec. The angular distance is maximum at $z \sim 1.6$, where 0.02 arcsec corresponds to 172 pc. This size is the low end of the ordinary dwarf spheroidal galaxies.

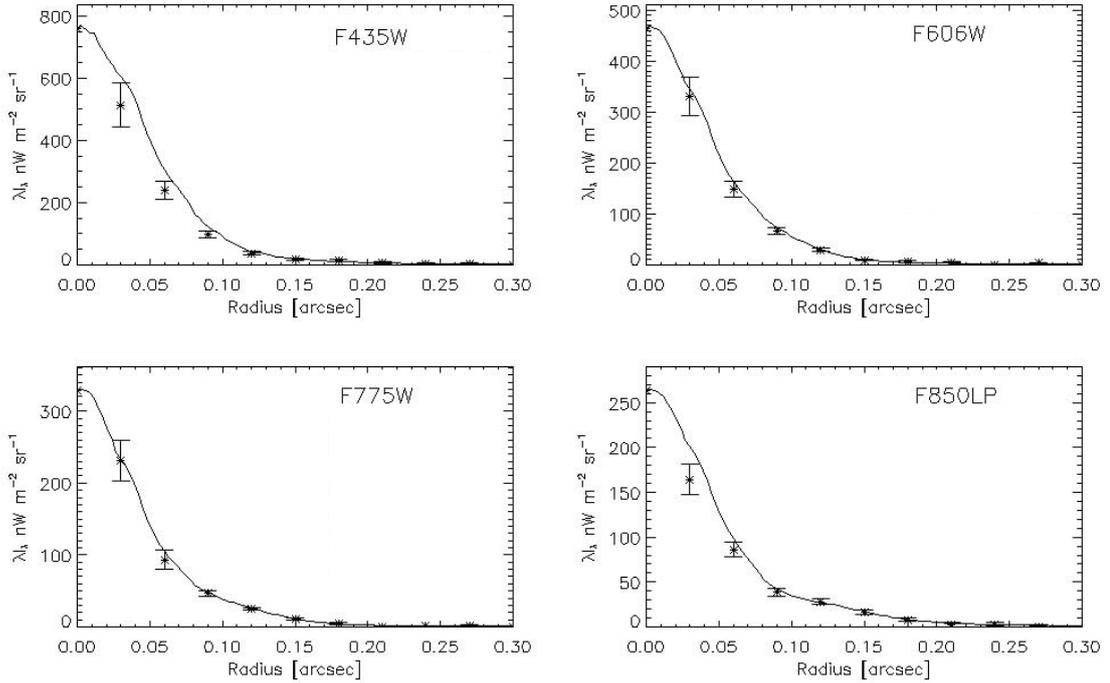

Fig. 9 Radial profiles of bright FCO, UDF8942. Solid lines indicate PSFs obtained from radial profiles of stars.

The spectral energy distribution (SED) is key to our understanding of the nature of FCOs. Since the fluxes of only four optical bands are presented in the UDF catalog[86], M&T used the UVUDF catalog[89], which provides fluxes for three ultraviolet (UV) and four infrared bands, alongside the optical bands. Fig. 10 presents the typical SEDs of four bright FCOs. The SEDs of the other bright FCOs show a similar feature. SEDs in the optical bands are featureless, exhibiting a power-law behavior, $\lambda I_\lambda \propto \lambda^{-1}$, which is typically found in starburst galaxies and AGN. In contrast, excess structures over the power law are found in the infrared bands. Interestingly, SEDs exhibit a drastic change at the borderline at 1 $\mu$m for almost all bright FCOs. This feature is very peculiar and has not been observed for the ordinary galaxies, AGNs or quasars.

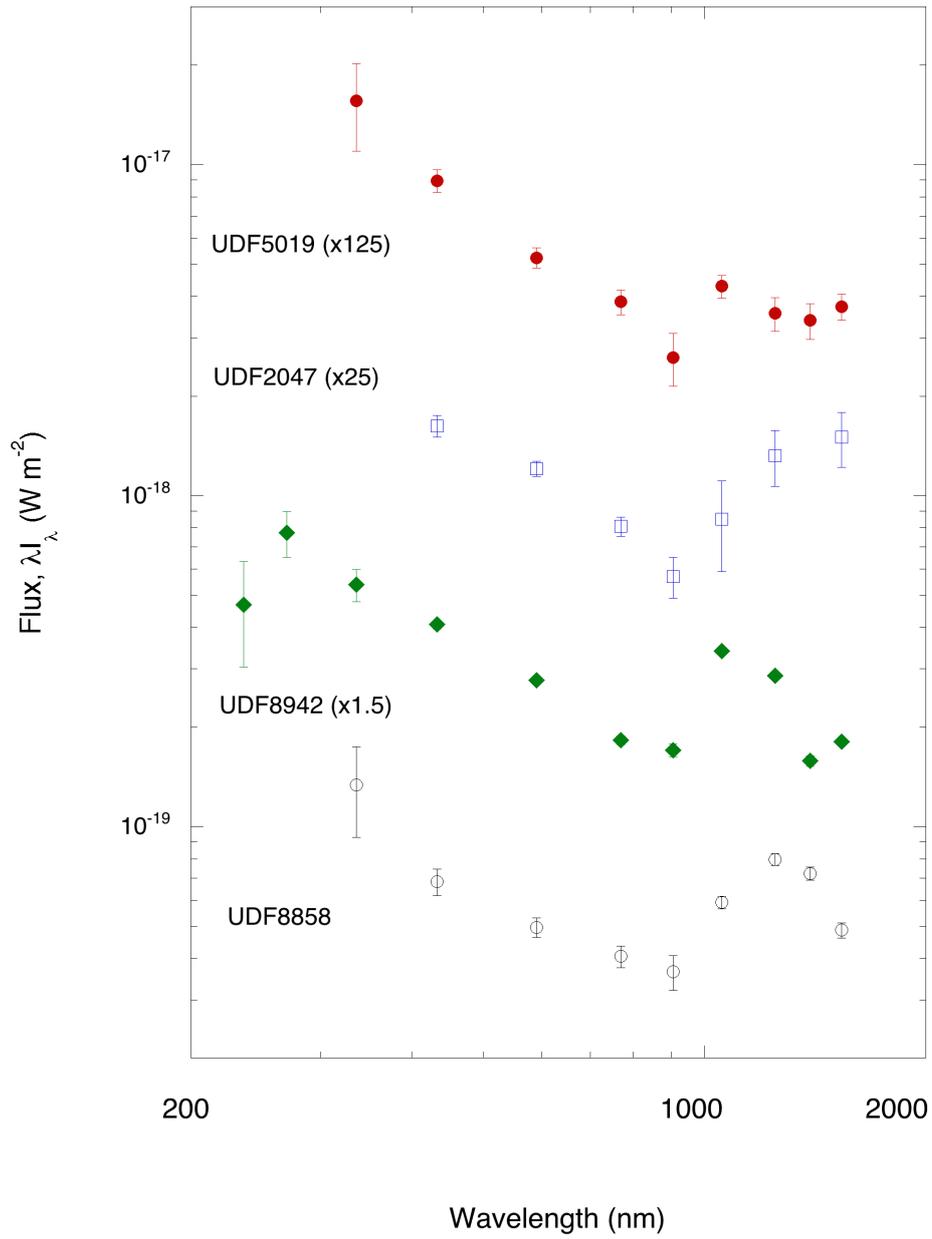

Fig. 10 Typical spectral energy distributions (SEDs) of four bright FCOs.

Rafelski et al. (2015)[89] used two evolution models of the galaxy to obtain the photometric redshifts of the UDF objects; however, uncertainties of the obtained redshifts for FCOs are considerable compared with those for ordinary galaxies to be used for further analysis. The fact that SED change at 1μm is common for FCOs suggests that redshifts of FCOs are almost the same; alternatively, FCOs are non-cosmological objects, i.e., very low redshift objects.

5. What are FCOs?

$\gamma$-ray observations are important for studying EBL, because $\gamma$-rays are absorbed by collisions with background photons. Biteau and Williams (2015)[90] analyzed all previous data of TeV-$\gamma$ observations with redshifts of lower than 0.6, and concluded that the obtained EBL was consistent with the integrated light of the known galaxies. The Fermi-LAT Collaboration (2018)[91] searched for EBL at high redshifts with lower energy $\gamma$-rays, and found that the obtained EBL was consistent with the evolution models of galaxies for a redshift range from 0.2 to 3.

These $\gamma$-ray observations imply that the origin of the excess brightness observed at XDF must be at the low redshift, at most 0.1. Assuming that FCOs are objects responsible for excess brightness and fluctuation, M&T examined the properties of FCOs.

M&T postulated that the surface number density of FCOs follows $10^{1.2m}$ toward the faint end, and spatial number density, $n(r)$, follows $r^3$ with cutoff at $z_{max}$ and $r_{max}$. In this case, the mass conservation law requires the existence of faded FCOs in our vicinity, with the total number of FCOs amounting to twice the active FCOs. M&T also attributed the mass of FCOs to the missing baryons (half of the baryons), because no other sources were available. Furthermore, for convenience, they assumed the same luminosity and mass for FCOs.

Since the total number of FCOs in the active phase was earlier obtained in a fluctuation analysis, they found the following properties of FCOs;

$L = 1.0 \times 10^3 \, L_{sun}(z_{max}/0.1)^2,$
$M = 3.2 \times 10^2 M_{sun}(z_{max}/0.1)^3,$
$M/L = 0.32 \, M_{sun}/L_{sun}(z_{max}/0.1).$

In the case of $z_{max} \sim 0.1$, the spatial number density, including faded FCOs, reaches $8.8 \times 10^6$ (Mpc)$^{-3}$. The distance to the detected FCOs ($\sim 30$ AB mag), for which M&T measured the radial profiles and SEDs is 48 Mpc, and the upper limit of the effective radius of 0.02 arcsec corresponds to 4.7 pc. The mass-to-luminosity ratio of FCOs is considerably lower than those of ordinary galaxies, and FCOs are much more compact than an ordinary stellar system with the same mass. These results indicate that the energy source of FCOs cannot be of a stellar or nuclear origin. They also suggest that FCOs could be powered by gravitational energy associated with black holes. FCOs may be considered to be mini-quasars since the mass and luminosity scales of FCOs are much lower than those of quasars.

A good correlation between the near-infrared and X-ray backgrounds supports the mini-quasar hypothesis. Cappelluti et al. (2017)[71] found that the contribution of the unknown population in the X-ray background is < $7 \times 10^{-13}$ erg cm$^{-2}$deg$^{-2}$, or < $2.3 \times 10^{-3}$ nWm$^{-2}$sr$^{-1}$, which corresponds to ~ $10^{-4}$ of the lower limit of the excess optical background. The ratio is very low compared with that of ordinary quasars, whose X-ray emission amounts to ~ 10% of the optical emission. This suggests that the X-ray emission mechanism of FCOs may not be so efficient compared with that of the quasars.

FCOs produce not only optical and near-infrared backgrounds, but also the X-ray background, and are very populous. Even in our vicinity, there exist numerous faded FCOs, which suggests that FCOs may be the origin of the gravitational waves.[80]

6. Discussion: future prospects

Fluctuation analysis has revealed that the sky is too bright to be explained with ILG free from the subtraction problem of ZL. However, detected excess brightness has a lower limit with the observed wavelength range being very narrow. The reliable and accurate SED of the excess brightness from UV to near-infrared wavelengths is vital. The most suitable mission for this purpose is observation outside zodiacal clouds. A simple photometer that covers a wide wavelength range from UV to near-infrared wavelengths will provide crucial data for investigating the structure and evolution of the universe in the recent epoch. EXo-Zodiacal Infrared Telescope (EXZIT)[92] now under consideration at Institute of Space and Astronautical Science (ISAS), Japan Aerospace Exploration Agency (JAXA), is a very promising mission.

As a candidate object responsible for excess brightness and fluctuation, M&T proposed FCOs. To confirm this hypothesis, the surface number density/magnitude relationship (Fig. 8) with fainter magnitudes must be examined. NASA's James Webb Space Telescope (JWST), which will be launched in 2021 will play a decisive role in this work. JWST has an aperture size of 6.5 m and an imaging and spectroscopic capability from 0.7 µm to a mid-infrared region. The detection limits at optical and near-infrared wavelengths are 2 ~ 3 magnitude better than those of the XDF observation, which clarify the separation of FCOs from ordinary galaxies in Fig. 8. The angular resolution of JWST is three-times as good as that of HST, which may allow us to resolve the spatial structure of FCOs. JWST's spectroscopic observations will provide information on the emission mechanisms and redshifts.

As for ground-based observations, the Thirty Meter Telescope (TMT) is very powerful. An excellent angular resolution of 0.01 arcsec and spectroscopic capability will be crucial to study FCOs.

We expect these observations to reveal the energy sources and emission mechanism, as well as the evolutional history of FCOs.

7. Conclusions

Recent observations of the optical and near-infrared background light were reviewed. Based on the fluctuation analysis of the XDF images, M&T showed that the blank sky has a structure down to 0.2 arcsec; they also verified the existence of fairly bright excess EBL free from the ZL subtraction problem. These findings require the existence of unknown celestial objects that are densely distributed across the sky.

M&T discovered FCOs, a new population, as a candidate for unknown objects. FCOs are very compact and have peculiar SEDs. Recent γ-ray observations indicate that FCOs are not high-$z$ objects, but are located in our vicinity ($z < 0.1$). If FCOs are responsible for excess brightness and fluctuation, their mass must be attributed to missing baryons; in other words, FCOs are the most populous objects in the universe. Furthermore, a very low *M/L* indicates that the energy source of FCOs could be gravitational energy associated with black holes.

The optical and near-infrared EBL and FCOs are now important observational targets. Further observations will shed light on the universe during the recent epoch.